\documentclass{mem}
\usepackage{natbib}\usepackage{txfonts}\usepackage{balance}
\usepackage{graphicx}
\usepackage[a4paper]{hyperref}
\idline{79}{1}
\begin{document}
\def\teff{$T\rm_{eff }$}
\def\kms{$\mathrm {km s}^{-1}$}

\title{
Testing adiabatic expansion of shocks in parsec-scale jets by dual-frequency VLBI experiments 
}

   \subtitle{}

\author{
A.B. \,Pushkarev\inst{1,2,3},
Y.Y. \,Kovalev\inst{1,4}
\and A.P. \,Lobanov\inst{1}
          }

  \offprints{A.B. Pushkarev}

\institute{
Max-Planck-Institut f\"ur Radioastronomie,
Auf dem H\"ugel 69, 53123 Bonn, Germany
\and
Crimean Astrophysical Observatory, Nauchnyj, Crimea, Ukraine
\and
Pulkovo Observatory, Pulkovskoe Chaussee 65/1, St. Petersburg 196140, Russia
\and 
Astro Space Center of Lebedev Physical institute, Profsoyuznaya 84/32, Moscow 117997, Russia 
\email{apushkarev@mpifr-bonn.mpg.de}
}

\authorrunning{Pushkarev }

\titlerunning{Adiabatic expansion of shocks in parsec-scale jets}

\abstract{ We present results of simultaneous dual-frequency (2 GHz
and 8 GHz) very long baseline interferometry (VLBI) observations of 12
active galactic nuclei with prominent jets.  Spectral properties of
the jets and evolution of their brightness temperature are
discussed. Measured sizes and brightness temperatures of VLBI features
are found to be consistent with emission from relativistic shocks
dominated by adiabatic losses. Physical scenarios with different
magnetic field orientation in the jets are discussed.
\keywords{galaxies: active -- galaxies: jets -- radio continuum:
galaxies} }

\maketitle{}

\section{Observations and Data Reduction}

Observations used for this work were made simultaneously at 2.3 and
8.6~GHz, with participation of all ten VLBA\footnote{Very Long
Baseline Array of the National Radio Astronomy Observatory (NRAO), Socorro
NM, USA} antennas and up to nine geodetic and EVN\footnote{European
VLBI Network} stations, in the framework of seven sessions of RDV
(Research \& Development~-- VLBA) observations started in 1994 under
coordination of the NASA and the NRAO and aimed at observations of compact
extragalactic radio sources. Four intermediate frequencies of 8~MHz
wide each were recorded making up a total 32~MHz bandwidth.
The data were correlated at the VLBA correlator in Socorro,
with a 4~sec integration time, and were obtained by us from the open
the NRAO archive and then calibrated using the NRAO Astronomical Image
Processing System (AIPS).
Phase corrections for residual delays and delay rates were done using
the task FRING.
Self-calibration, hybrid mapping, and model fitting were performed in
DIFMAP \citep{Shepherd94}. In the model fitting, we used a minimum number
of circular Gaussian components that was reproducing adequately the
observed interferometric visibilities.

\section{Results}

\begin{figure}[t!]
\resizebox{\hsize}{!}{\includegraphics[clip=true]{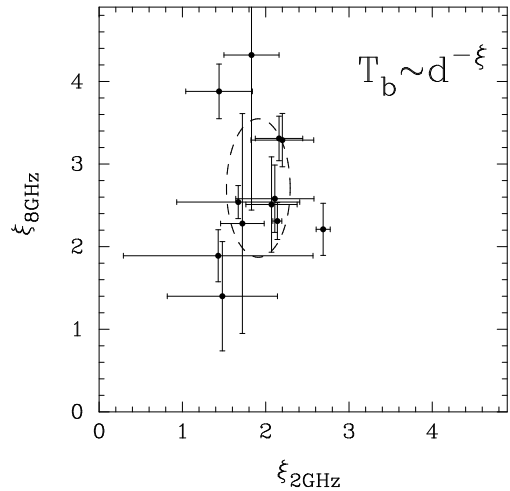}}
\caption{\footnotesize Distribution of power-law indices of brightness
temperature gradients with jet component size at 2.3 and 8.6~GHz.  The
dashed ellipse represents a one-sigma error area of $\xi$-index
distributions.  }
\label{xi-index}
\end{figure}

We selected 12 active galactic nuclei (out of 222 observed in the
RDV31-37 sessions) with prominent jets having at least 3 jet
components detected at both frequencies. For these objects, we
analyzed brightness temperature evolution as a function of (i)
distance to VLBI core, $r$, (ii) size of jet component, $d$. For all
components, flux densities, sizes, and relative positions to the core
have been obtained from model fitting of the self-calibrated data.
Evolution of $T_\mathrm{b}$ can be described with power law functions
$T_{\rm b}\propto r^{-k}$ and $T_{\rm b}\propto d^{-\xi}$. The power
law index $k$ varies between 1.2 and 3.6, with the average value of
$\overline{k}_{\,\rm{8\,GHz}}\approx\overline{k}_{\,\rm{2\,GHz}}\approx2$.
The power law index $\xi$ varies between 1.4 and 4.3, with the average
values of $\overline{\xi}_{\,\rm{8\,GHz}}=2.7$ and
$\overline{\xi}_{\,\rm{2\,GHz}}=1.9$. Distributions of the fitted
power indices $k$ and $\xi$ are shown in Figs.~\ref{k-index},
\ref{xi-index}.

\begin{figure}[t!]
\resizebox{\hsize}{!}{\includegraphics[clip=true]{pushkarev_f1.ps}}
\caption{\footnotesize Distribution of power-law indices of brightness
temperature gradients with distance to the core at 2.3 and 8.6~GHz.
The dashed ellipse represents a one-sigma error are of $k$-index
distributions.  }
\label{k-index}
\end{figure}

The dependence of $T_{\rm b}\propto d^{-\xi}$ that also includes VLBI
core components may be used for testing the shock-in-jet model
\citep{Marscher90}, with the component emission dominated by adiabatic
energy losses in relativistic shocks developing in a flow with a power
law particle energy distribution $N(E)\,{\rm d}\,E\propto E^{-s}\,{\rm
d}\,E$ and the magnetic field $B\propto d^{-a}$. Here, $d$ is the
transverse jet size and $a$ describes the orientation of the magnetic
field ($a=1$ for a transverse field and $a=2$ for a longitudinal
field). Assuming the Doppler factor changing weakly along the jet, the
brightness temperature of each jet component, $T_{\rm b,\,jet}$, can
be related to the measured brightness temperature of the core, $T_{\rm
b,\,core}$, as $T_{\rm b,\,jet}=T_{\rm b,\,core}(d_{\rm jet}/d_{\rm
core})^{-\xi}$, where $d$ represents the measured sizes of the core
and jet components, and $\xi=[2(2s+1)+3a(s+1)]/6$
\citep{Lobanov00}. For the spectral index $\alpha=(1-s)/2$
($S\propto\nu^\alpha$), we obtain $\xi=a+1-\alpha(a+4/3)$. With $\xi$
determined from the data, we can test the shock-in-jet model by (i)
comparing the jet brightness temperatures predicted by the model with
those from the data; (ii) choosing the appropriate pair ($\alpha_{\rm
mod}$, $a=1$) or ($\alpha_{\rm mod}$, $a=2$) by comparing with
observed jet spectral indices obtained after applying core shift
correction \citep{Kovalev08}; (iii) using VLBA polarization observations 
from which the magnetic field orientation is determined directly.

\begin{figure}[t!]
\resizebox{\hsize}{!}{\includegraphics[clip=true,angle=-90]{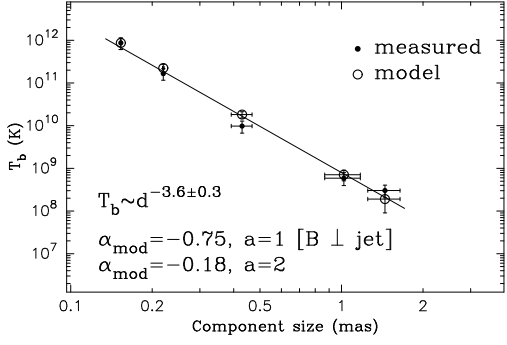}}
\caption{
\footnotesize
Brightness temperature as a function of jet components for BL Lacertae object $1823+568$.
}
\label{Tb_1823}
\end{figure}

\begin{figure}[t!]
\resizebox{\hsize}{!}{\includegraphics[clip=true]{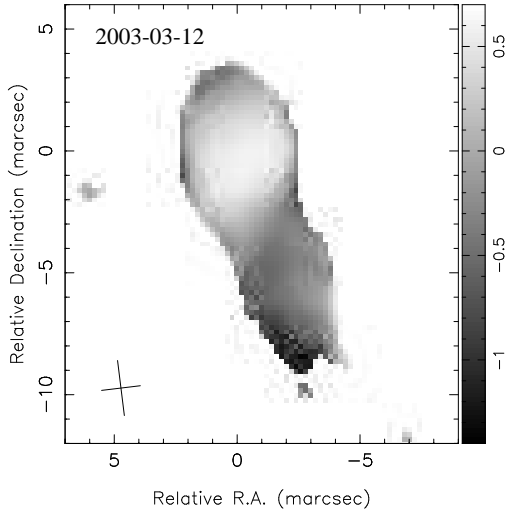}}
\caption{
\footnotesize
Distribution of spectral index $\alpha_{2-8\,\rm GHz}$ in $1823+568$.
}
\label{alpha_1823}
\end{figure}

\begin{figure}[t!]
\resizebox{\hsize}{!}{\includegraphics[clip=true,angle=-90]{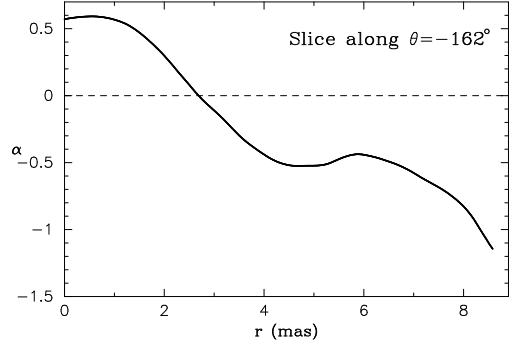}}
\caption{
\footnotesize
Profile of spectral index $\alpha_{2-8\,\rm GHz}$ along the jet
direction in $1823+568$.
}
\label{slice_1823}
\end{figure}

\begin{figure}[t!]
\resizebox{\hsize}{!}{\includegraphics[clip=true]{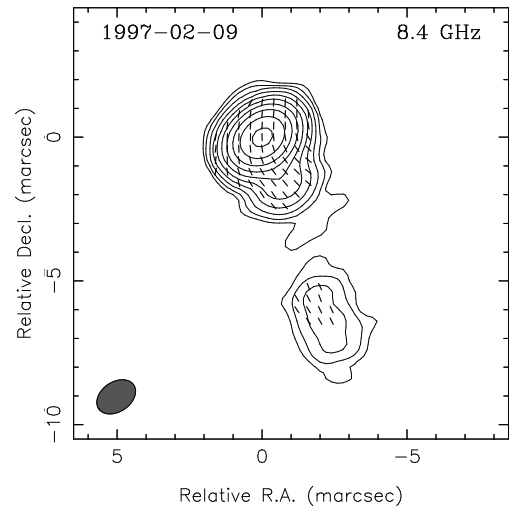}}
\caption{
\footnotesize
Total intensity image of $1823+568$ at 8.4~GHz with the polarization vectors superimposed.
}
\label{B-field_1823}
\end{figure}

\begin{figure}[t!]
\resizebox{\hsize}{!}{\includegraphics[clip=true,angle=-90]{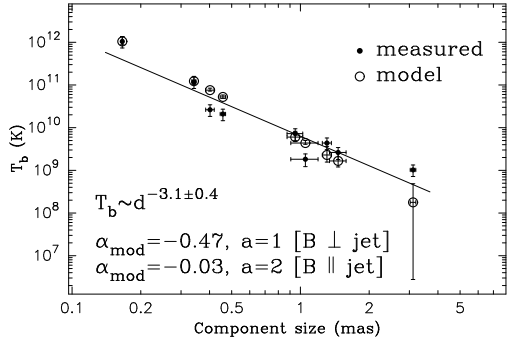}}
\caption{
\footnotesize
Brightness temperature as a function of jet components for quasar CTA~102.
}
\label{Tb_CTA102}

\end{figure}

\begin{figure}[t!]
\resizebox{\hsize}{!}{\includegraphics[clip=true]{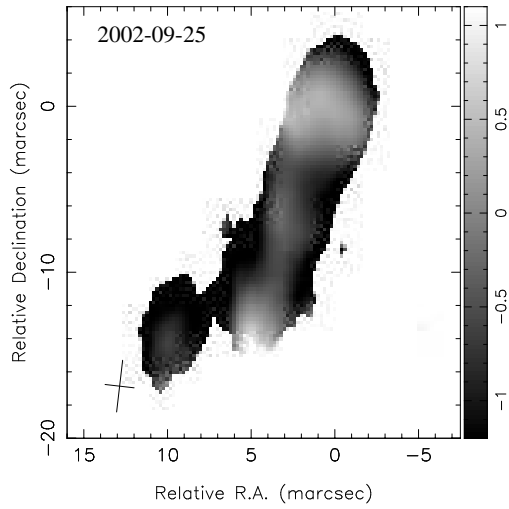}}
\caption{
\footnotesize
Spectral index $\alpha_{2-8\,\rm GHz}$ distribution map of CTA~102.
}
\label{alpha_CTA102}
\end{figure}

\begin{figure}[t!]
\resizebox{\hsize}{!}{\includegraphics[clip=true,angle=-90]{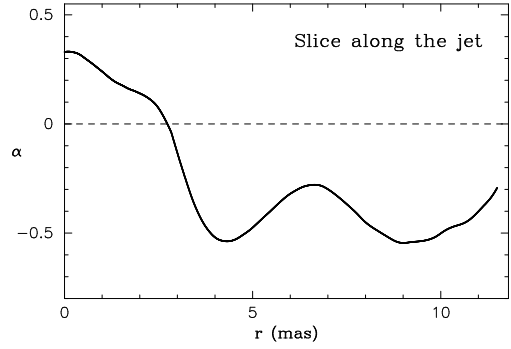}}
\caption{
\footnotesize
Slice of spectral index $\alpha_{2-8\,\rm GHz}$ distribution map along the jet
direction in CTA~102.
}
\label{slice_CTA102}
\end{figure}

\begin{figure}[h!]
\resizebox{\hsize}{!}{\includegraphics[clip=true]{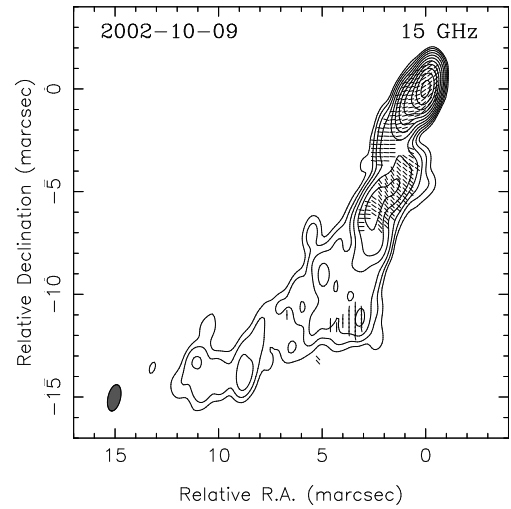}}
\caption{
\footnotesize
Total intensity image of CTA~102 at 15~GHz with the polarization vectors superimposed.
}
\label{B-field_CTA102}
\end{figure}

Results of such an analysis are presented in
Figs.~\ref{Tb_1823}-\ref{B-field_1823} for the BL Lac object
$1823+568$ and in Figs.~\ref{Tb_CTA102}-\ref{B-field_CTA102} for the
quasar CTA~102. In both cases, the model brightness temperatures agree
with the measured values. Our conclusion about the transverse B-field
in $1823+568$ (Fig. \ref{Tb_1823}) is supported by the polarization
map at 8.4~GHz in Fig.~\ref{B-field_1823} \citep{Pushkarev08}.  In the
quasar CTA~102, both transverse 
and longitudinal magnetic field orientations are consistent with the model
(Fig. \ref{Tb_CTA102}). This is confirmed by MOJAVE \citep{Lister05}
observations of this source (Fig. \ref{B-field_CTA102}). In the
innermost part of the jet, the magnetic field is predominantly
transverse, and it becomes longitudinal at larger distances,
implying dissipation of shocks and creation of polarization sheath
around the jet. It should also be noted that the observed brightness
temperature of the largest ($d=3.1$~mas) jet component located at
$\approx12$~mas from the core is higher than the model brightness
temperature ($T_{\rm \,b}^{\rm obs}/T_{\rm \,b}^{\rm model}\approx6$).
This can be explained by an interaction with the
ambient medium, also seen in the spectral index map
(Fig.~\ref{alpha_CTA102}).

\section{Conclusions}
Measured sizes and brightness temperatures of
VLBI components in BL Lac object $1823+568$ and quasar CTA~102 are 
consistent with emission from relativistic
shocks dominated by adiabatic 
losses.

\begin{acknowledgements}
This work is based on the analysis of global VLBI observations
including VLBA, the raw data for which were provided to us by the 
open NRAO archive. The National Radio Astronomy Observatory is a 
facility of the National Science Foundation operated under 
cooperative agreement by Associated Universities, Inc. 
This research has made use of data from the MOJAVE \citep{Lister05} 
and 2cm Survey \citep{Kellermann04} program.
Y.~Y.~Kovalev is a Research Fellow of the Alexander von Humboldt Foundation.
\end{acknowledgements}



\bibliographystyle{aa}

\end{document}